# Polar Rectification Effect in Electro-Fatigued SrTiO$_3$ Based Junctions

Xueli Xu,[1,2,8] Hui Zhang,[3,8] Zhicheng Zhong,[4,*] Ranran zhang,[1] Lihua Yin,[5] Yuping Sun,[5] Haoliang Huang,[6] Yalin Lu,[6] Yi Lu,[7] Chun Zhou,[1] Zongwei Ma,[1] Lei Shen,[1,2] Junsong Wang,[1] Jiandong Guo,[3] Jirong Sun,[3,*] Zhigao Sheng[1,9,*]

[1]Anhui Province Key Laboratory of Condensed Matter Physics at Extreme Conditions, High Magnetic Field Laboratory of the Chinese Academy of Science, Hefei, Anhui 230031, China.

[2]University of Science and Technology of China, Hefei, 230026, China

[3]Beijing National Laboratory for Condensed Matter Physics, Institute of Physics, Chinese Academy of Sciences, Beijing 100190, China

[4]Key Laboratory of Magnetic Materials and Devices, Ningbo Institute of Materials Technology and Engineering Chinese Academy of Sciences Ningbo 315201, China

[5]Key Laboratory of Materials Physics, Institute of Solid State Physics, Chinese Academy of Sciences, Hefei 230031, China

[6]Anhui Laboratory of Advanced Photon Science and Technology, National Synchrotron Radiation Laboratory, University of Science and Technology of China, Hefei 230026, P. R. China

[7]Institute for Theoretical Physics, Heidelberg University, Philosophenweg 19, 69120 Heidelberg, Germany

[8]These authors contributed equally

[9]Lead Contact

*Correspondence
zhigaosheng@hmfl.ac.cn (Z. Sheng)
zhong@nimte.ac.cn (Z. Zhong)
jrsun@iphy.ac.cn (J. Sun)

**SUMMARY**

Rectifying semiconductor junctions are crucial to electronic devices. They convert alternating current into direct one by allowing unidirectional charge flows. In analogy to the current-flow rectification for itinerary electrons, here, a polar rectification that based on the localized oxygen vacancies (OVs) in a Ti/fatigued-SrTiO$_3$ (fSTO) Schottky junction is first demonstrated. The fSTO with OVs is produced by an electro-degradation process. The different movability of localized OVs and itinerary electrons in the fSTO yield a unidirectional electric polarization at the interface of the junction under the coaction of external and built-in electric fields. Moreover, the fSTO displays a pre-ferroelectric state located between paraelectric and ferroelectric phases. The pre-ferroelectric state has three sub-states and can be easily driven into a ferroelectric state by external electric field. These observations open up opportunities for potential polar devices and may underpin many useful polar-triggered electronic phenomena.

**Progress and potential**

Charges in a matter can be either itinerant or localized. The current rectifier devices based on itinerant charge have been realized at the end of the19th century and become a building block in current electronics. The polar rectifier device based on localized charges, which holding great potential applications in future electronics, has not been realized yet. In this manuscript, we present a first demonstration of polar rectification based on the localized oxygen vacancies (OVs) in a Ti/fatigued-SrTiO$_3$ (fSTO) Schottky junction. The fSTO with OVs is produced by an electro-degradation process. The different movability of localized OVs and itinerary electrons in the fSTO yield a unidirectional electric polarization at the interface of the junction under the coaction of external and built-in electric fields. Our observations open up opportunities for potential polar devices and may underpin many useful polar-triggered electronic phenomena.

**INTRODUCTION**

From the perspective of movability, charges can be divided into two types: itinerant and localized.[1] The movement of itinerant charge carriers generates a current under an electric field. Metals or ohmic contacted heterojunctions exhibit symmetric *J-V* curves where the intensity of current *J* is independent of the direction of electrical field V (Figure 1A). For heterojunctions with a non-ohmic contact, a rectifying effect with asymmetric *J-V* relation is expected,[2] as schematically shown in Figure 1B. The rectification of itinerant charge carriers has become a key signal-processing tool in current electronics, ranging from power supplies to high frequency detectors, smart phones, computers, TV-sets and wireless communications. The localized charges, on the other hand, can generate electric polarization when placed in an electric field. In general, the relationship between the polarization strength (*P*) and *V* is symmetric for either a single dielectric material or a non-ohmic contacted heterointerface (Figure 1C). Finding an asymmetric *P-V* characteristic, i.e., polar rectification, is a natural expectation, and it may have potential applications in the field of switchable rectifier, energy storage, and computation.

In this context, the manipulation of remnant polarization, especially at the surface/interface, is of great importance for the creation of polar rectification. Recently, it has been shown that polar switching of ferroelectric thin films is sensitive to the electric boundary conditions at the surface/interface, including space charge generated from band bending or charged states formed from defects/surface adsorbates.[3,4] The electronic asymmetry at the ferroelectric surface/interface, especially space charge accumulated near the metal-ferroelectric interfaces, can often generate a displacement of the *P-V* hysteresis loops, particularly along the electric field axis.[3,4] Such displacement is a biased effect with negligible on/off ratio, and the polar rectification phenomenon has not been reported so far. However, the idea related to the localized charge manipulation near the metal-dielectric interfaces may provide a possible route for the realization of polar rectification.

Oxygen vacancies (OVs) in oxides can act as either mobile or localized donors, and their behavior at the metal-oxide interface can be controlled electrically.[5] As a typical transition metal oxide with high dielectric constant, low dielectric loss, and good thermal stability, $SrTiO_3$ (STO) is widely used in inverter capacitors, resistance switching random access memories, and other electronic devices.[6,7] Considerable research efforts have been devoted to the OV-induced structural, electric, magnetic, and ferroelectric properties in STO single crystals and metal-STO heterojunctions.[6-13] In addition, STO was identified as an incipient ferroelectric material, which approaches but does not cross the ferroelectric phase transition due to quantum fluctuation.[14,15] Introducing OVs can suppress the quantum fluctuation and stabilize the ferroelectric phase,[16] yet unidirectional electric polarization has not been reported. Here, by repeatedly applying the electric field to Ti/STO heterojunction, an electrical-fatigued STO (fSTO) crystal with OV-electron pairs is created. By controlling the movement of OV-electron pairs in the depletion layer of Ti/fSTO junction, a polarization rectification effect is realized. Further structural, dielectric, Raman, and nonlinear optical measurements show that such fSTO resides in the crossover regime between the pristine quantum paraelectric state and the ferroelectric state. A small external electric field (<1.2 kV/cm) can drive this pre-ferroelectric STO crystal into ferroelectric phase within the temperature range from 50 K to 170 K.

**RESULTS AND DISCUSSION**

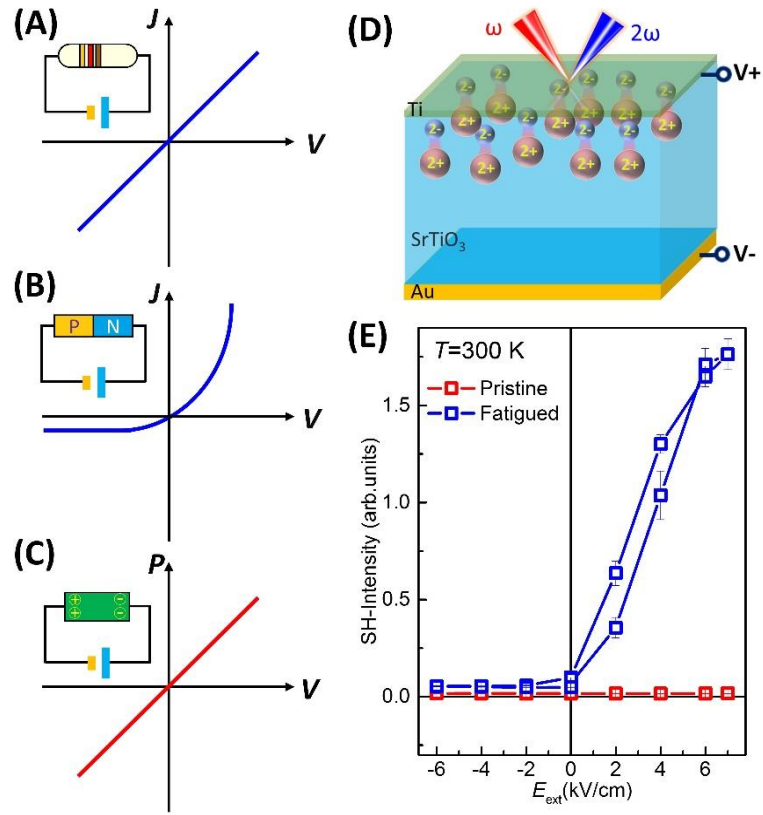

**Figure 1. Schematic images of *J-V* and *P-V* relationships for typical electronic devices and SHG experimental results of Ti/STO junctions.**

(A) The J-V relationship of a resistor.
(B) The J-V relationship of a p-n junction.
(C) The P-V relationship of a dielectric material.
(D) Schematic image for a Ti/fSTO/Au heterojunction and SHG measurements for the electric polarization. The red and blue balls represent the oxygen vacancies and electrons, respectively.
(E) Electric field dependent SH-intensity for both pristine and fatigued Ti/fSTO junction measured at $T = 300$ K. The arrows represent the electric field sweeping direction.

The structure of the Ti/STO heterojunctions is sketched in Figure 1D. A (001)-orientated STO single crystal (5×5×0.5 mm$^3$) was covered by a 6 nm-thick Ti layer (anode) on the surface and a 100 nm-thick Au layer (cathode) on the back. Both pristine-STO (pSTO) and fSTO samples were studied, with the latter produced by sweeping electric field from 0 to 10 kV/cm up and down more than ten times on the Ti/STO junction. Before, during, and after the fatigue process, the electric polarization in the Ti/fSTO junction is measured by the second harmonic generation (SHG) technique, which is highly sensitive to symmetry breaking and often used in probing electric polarization in ferroelectric crystals and thin films (Supplementary Figure S1).[17-19] Figure 1E shows the SH intensity as a function of external electric field ($E_{ext}$) for the Ti/fSTO heterojunction. As sweeping $E_{ext}$ from negative to positive values, the SH output signal for the Ti/pSTO junction is negligible and remains constant in the whole measured field range (red line in Figure 1E). In contrast, for the Ti/fSTO case, it is apparent to find an asymmetric SH-$E_{ext}$ loop as $E_{ext}$ sweeps (blue line in Figure 1E). When a negative $E_{ext}$ is applied, i.e., Ti(-)/STO(+), the SH output signal is small and remains constant, similar to that of the pristine case. For positive $E_{ext}$, on the other hand, the SH intensity increases monotonously with increasing $E_{ext}$. The ratio of SH signal for $E_{ext}$ = +6 and -6 kV/cm is around 32, which can serve as the on/off ratio of this polar rectification junction. Comparing the SH-$E_{ext}$ results of both pristine and electro-fatigued STO junction, we conclude that the fatigue process of STO crystal plays a key role in realizing polar rectification.

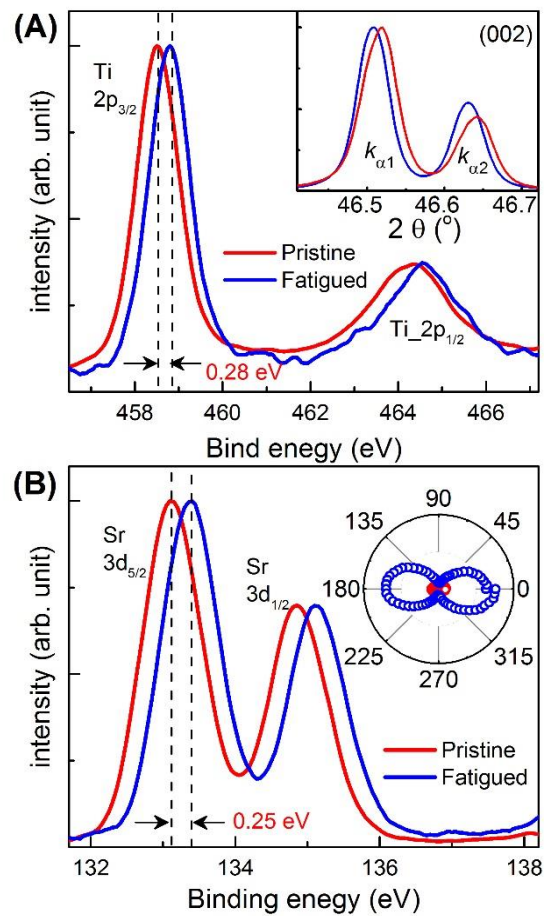

**Figure 2. XPS spectra of for both pSTO and fSTO crystal.**
(A) The XPS spectra of Ti in STO crystals. Inset of (A) is the XRD results of (002) peak for both pSTO and fSTO crystal.
(B) The XPS spectra of Sr in STO crystals. Inset of B) is the polar plots of the SH intensity versus

the polarization angle of output light for $E_{ext}$ = 0 V/cm of pSTO and $E_{ext}$ = 0 V/cm, 10 kV/cm of fSTO.

As mentioned above, the fatigue process of STO crystal was done by sweeping large $E_{ext}$ up and down repeatedly in the Ti/STO junction. This is similar to the processes adopted in previous works,[10,11,20] where electric-field-fatigued STO was reported. It was also found that the Ti/STO interface can act as an oxygen getter. The work function ($\Phi_M$) of Ti is 4.33 eV. The band gap, electron affinity, and charge neutrality level for SrTiO$_3$ are 3.3 eV, 3.9 eV, and 4.6 eV, respectively.[21-23] The bands lineup between the work function of metal and the charge neutrality level of STO, which gives a Schottky barrier height (built-in potential) of -0.27 eV (band bending down) for Ti/STO (Supplementary Figure S2). The electric field in the interface of depletion layer ($E_{in}$) is calculated at ~ 45.8 kV/cm based on the sharp Schottky diode model (Supplementary Figure S2), which is much larger than the $E_{ext}$ applied here. When the applied $E_{ext}$ is opposite to $E_{in}$, i.e., Ti(-)/STO(+), no OVs can be generated as $E_{ext}$ is much smaller than $E_{in}$. For the Ti(+)/STO(-) case, however, the OVs can be continually generated near the Ti/STO interface when $E_{ext}$ exceeds a critical value ($E_c$).[11] For the Ti/pSTO(001) junction, the $E_c$ is around 8 kV/cm. In this work, with sweeping $E_{ext}$ up and down between 0 and 10 kV/cm (>$E_c$), OVs are generated and migrate into the STO. After 10 electric cycles, many remnant OVs were produced near the Ti/STO interface and thus an electro-fatigued STO layer is created (Figure 1D). Due to the large intensity of $E_{in}$, most remnant OVs were located in the depletion layer of Ti/ STO junction. It was calculated that the width of depletion layer ($W_d$) increases from one hundred nanometers to several micrometers when $E_{ext}$ was swept from 0 to 10 kV/cm (Supplementary Figure S2). We emphasize that such electro-fatigue process can only be realized in the metal/STO junction with bending down of electron energy band. For the band bending up ($\Phi_M$ > 4.6 eV) cases, such as Au/STO, the remnant OVs and thus fSTO could not be produced due to the opposite direction of $E_{in}$, which has been confirmed by X-ray measurements (Supplementary Figure S3). These features are also consistent with previous reports, in which the oxygen atoms were proved to be able to transfer from the STO to the metal layer resulting in the formation of OVs in STO.[12,13,24]

In fSTO, the ejection of the neutral oxygen leaves a lattice deformation (OV) and two extra electrons. These electrons can stay in the OVs accompanied by strong relaxation of the neighborhood lattice (a polaron).[25] For simplicity, the OVs with attracted electrons are labeled as OV-electron charge pairs here. With an $E_{ext}$ applied, the displacement of electrons from the OVs in the fSTO would produce an electric polarization, which can be detected by the SHG measurement. In our experiment, the whole electro-fatigued process was monitored by the SH measurements (Supplementary Figure S4). It was found that the $E_{ext}$ = 10 kV/cm is enough to fatigue (001)-oriented STO crystal, while it is insufficient for (110) and (111) STOs ($E_c$ > 10 kV/cm), which indicates that the creation and migration of OVs is harder in the latter two crystalline directions (Supplementary Figure S5).

To further confirm the existence of OVs in the fSTO, X-ray photoelectron spectroscopy (XPS) was performed. Figure 2A and 2B show the high-resolution XPS spectra of Ti 2p and Sr 3d for the fatigued and pristine STO, respectively. These XPS data were calibrated by the C1s spectra (Supplementary Figure S5). Comparing with that of the pSTO, the XPS peaks of the fSTO shift to higher energy. Specifically, the peaks of Ti $2p_{1/2}$ and Ti $2p_{3/2}$ shift about 0.28 eV and the two peaks of Sr 3d shift about 0.25 eV. This indicates the existence of OVs in the STO.[26] In addition, X-ray diffraction (XRD) results show that the (002) diffraction peak of fSTO crystal shifts to lower 2θ angle (the inset of Figure 2A), which indicates the elongation of *c*-axis and therefore the existence of OVs in the whole near surface region.[10,11]

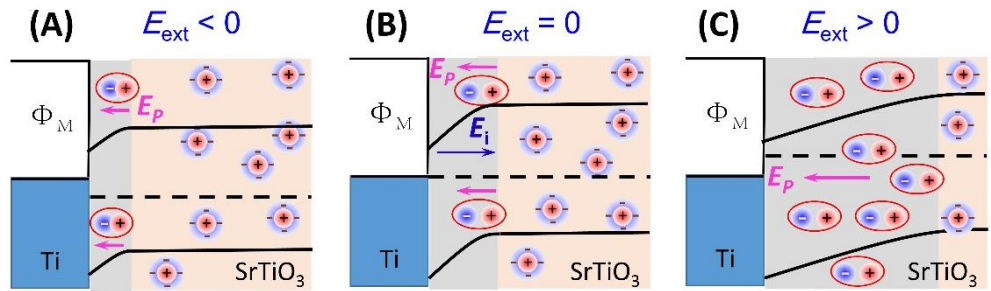

**Figure 3. The schematic illustration of energy band diagram of Ti/fSTO junctions.**
(A-C) presents the band structures with (A) $E_{ext} < 0$ V/cm, (B) $E_{ext} = 0$ V/cm and (C) $E_{ext} > 0$ V/cm. The work function ($\Phi_M$) of Ti is 4.33 eV, and the charge neutrality level of STO is 4.6 eV. The pink and blue arrows indicate the polarization direction of OV and electrons pairs and the electric field in the depletion layer of the junction.

When a conventional dielectric material, such as pSTO, is placed in an electric field, the positive and negative charges slightly shift from their average equilibrium positions in an opposite way. Nevertheless, the overall positive and negative charge centers in the material are still coincident. Then there is no remnant electric polarization exist, except minor field-induced charges on surface, which would contribute to the polarization as well as the SH output. The amount and the direction of those surficial charges are determined by both the intensity and the direction of $E_{ext}$. Hence, a symmetric $P$-$E_{ext}$ relationship can be obtained (Figure 1C). This is the case for Ti/pSTO junction (red curve in Figure 1E). However, the situation is different if there are excess charges, such as OVs. These excess OVs-electron pairs can be separated by $E_{ext}$, producing electric polarization as well as SH output, which has been confirmed by our experiments (blue curve in Figure 1E). It should be noted that the movability of positive OVs and negative electrons are completely different. While the electrons can be easily moved by the electric field, as a structural defect, the movement of OVs requires higher energy as proved by the fatigue process. The $E_{ext}$ values in our study are smaller than the highest electric field used for the fatigue process (10 kV/cm), which only drives the electrons around essentially static OVs.

Figure 3A-3C shows the schematic band diagram of Ti/fSTO Schottky junctions with different $E_{ext}$. When $E_{ext} = 0$, the band bending and the charge accumulation happens in the $W_d$ (~133.8 nm), accompanied with $E_{in}$ pointing from Ti to STO (Figure 3B). The OV-electron pairs in the depletion layer were separated due to the existence of $E_{in}$, which produce a small remnant polarization as well as weak SH signals (Figure 1E and blue circles in the inset of Figure 2B). When $E_{ext} < 0$, $W_d$ shrinks and the amount of OVs in the depletion layer is reduced. As the result, less polarization as well as weaker SH signals can be detected (Figure 1E). When $E_{ext} > 0$, as shown in Figure 3C, $W_d$ extends wider and the number of OV-electron pairs in the depletion layer increases much. At the same time, the $E_{ext}$ induced separation of OV-electron pairs gives rise to much larger polarization as well as SH signal. It was found that, comparing to the SH result obtained with $E_{ext} = 0$, a 7 kV/cm $E_{ext}$ causes 25 times increase of SH output (Figure 1E and the inset of Figure 2B), which corresponds to a 5-fold enhancement of $P$. Obviously, due to the low movability of OVs and the variation of $W_d$ in Ti/fSTO Schottky junction, an asymmetric SH-$E_{ext}$ relationship was obtained as shown in the Figure 1E. This polar rectification observed in fSTO is analogous to the current rectification in semiconductor heterojunctions. This electric field induced rectification effect of electrical polarization may have potential applications in the memory device, high voltage protection, and so on.

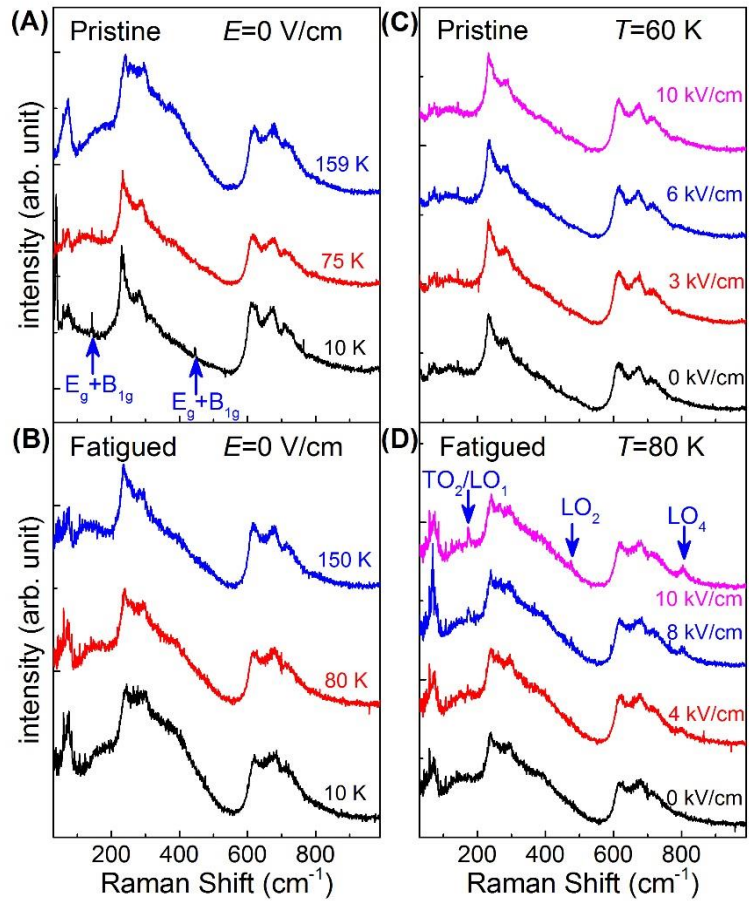

Figure 4. The Raman spectrum of pristine and fatigued STO.

(A) and (B) The Raman spectrum at different temperatures when $E = 0$ V/cm for pristine STO and fatigued STO, respectively.

(C) and (D) The Raman spectrum at different electric fields of pristine STO ($T = 80$ K) and fatigued STO ($T = 60$ K), respectively.

The blue arrows in (A) indicate the $E_{1g}+B_{1g}$ modes. The blue arrows in (D) indicate the TO/LO modes.

The OVs distort the lattice structure of STO crystal (inset of Figure 2A), which may also give rise to a transition of the electronic phases of STO crystal. To further clarify the exact phase state of fSTO, Raman spectra were measured at different temperatures and the typical results are shown in Figure 4. For pSTO, there is a cubic to antiferrodistortive (AFD) structure transition at $T_a \sim 105$ K and the Eg+B1g (143cm$^{-1}$, 446cm$^{-1}$) Raman modes emerge below $T_a$.[27] This has been confirmed by Figure 4A. For fSTO, however, the Raman spectra are almost temperature independent (Figure 4B). No Eg+B1g Raman modes are found down to 10 K, which implies a suppression of the AFD structure transition by OVs. More than that, no ferroelectric phase is formed, as implied by the absence of first-order LO phonons modes that has been identified as the signature of ferroelectric phase in perovskite oxides.[28] Figure 4C and 4D show the Raman spectra under $E_{ext}$ for both pristine and fatigued STO measured at $T = 60$ K and 80 K, respectively. It was found that the Raman spectra of pSTO are insensitive to electric field up to 10 kV/cm. Interestingly, a series of first-order LO phonon Raman modes, i.e., peaks at 173 (LO1/TO2), 475 (LO3), 803 cm$^{-1}$ (LO4), as well as the ferroelectric structural phase were induced in fSTO by external electric field.[28] It implies that, comparing with pSTO, the fSTO can be easily driven into the ferroelectric state. These observations indicate that the OVs in fSTO has suppressed AFD and driven the STO to the border of the FE phase. From this viewpoint, it is reasonable to deduce that the fSTO is in a pre-ferroelectric state.

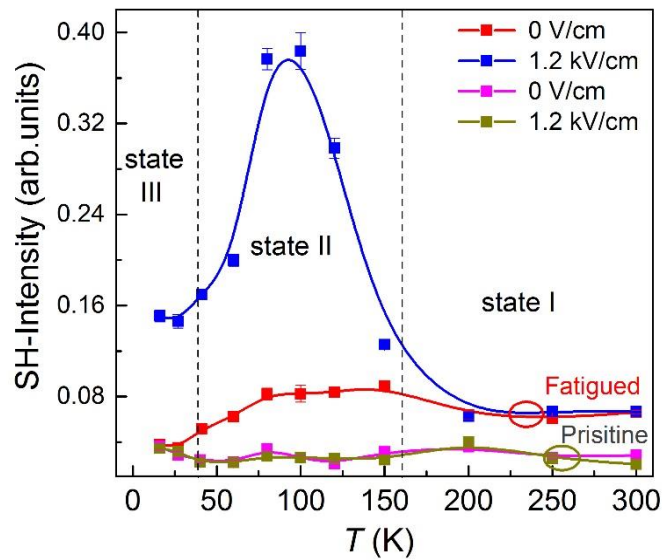

Figure 5. The temperature dependence of SH intensity for fatigued and pristine STO with $E_{ext}$

= 0 and 1.2 kV/cm respectively.

To further explore the pre-ferroelectric state in fSTO, the temperature dependent SH output has been studied. As depicted in Figure 5, the SH measurement of pSTO shows that there is almost no spontaneous polarization in whole measurement temperature range. As for the fatigued sample, there is an upturn of SH signal around 170 K and a drop of SH output below 40 K (red curve in Figure 5). This feature implies that there are three sub-states for the pre-ferroelectric STO crystal at different temperatures. We assign them as pre-ferroelectric state I ($T > 170$ K), II (40-170 K), and III (below 40 K), respectively. To further study the different pre-ferroelectric states of fatigue STO, the temperature dependent SH measurements with application of a small electric field ($E_{ext}$ = 1.2 kV/cm) were done (blue curve in Figure 5). For pre-ferroelectric state I, there is negligible effect on SH signal with such small $E_{ext}$. When the fSTO get into pre-ferroelectric state II, a large enhancement of SH output was found. Around 80 K, the SH is enlarged more than 5 times by a 1.2 kV/cm electric field. Such electric field enhancement remains below 40 K, which implies the application of $E_{ext}$ help suppress quantum fluctuation in the fSTO. The temperature dependence of pre-ferroelectric state might originate from the temperature dependence of ionization of OVs in fSTO. Based on the SH results, it seems that the OVs-electron pairs are more easily separated in the temperature range of 40-170 K, which may be attributed to the structural transition of fSTO crystal within this temperature range.

**Conclusion**

In conclusion, by utilizing electric field sweeping technique, the fSTO with OVs inside was created. Taking advantages of different movability between OVs and electrons in the depletion layer, a polar rectification was realized for the first time in a Ti/fSTO heterojunction. The temperature and electric field varied SH output, X-ray diffraction and Raman measurements, we identified that such STO owns a pre-ferroelectric state, which is just between quantum paraelectric and ferroelectric states. Moreover, it is interesting to find that the fSTO with pre-ferroelectric state is easily to be driven into ferroelectric state by an external electric field. These findings not only give further path to study the ferroelectric state of perovskite oxides, but also provide a novel electric rectification device that may be a building block in future electronics.

**EXPERIMENTAL PROCEDURES**

**Sample fabrication.** The Ti/STO/Au sandwiched samples for SH and dielectric measurements were obtained by depositing 6-nm-thick Ti layer on the top and 100 nm-thick Au layer on back side of a (001)-STO substrate (3 × 5 × 0.5 mm³, doubly polished), respectively. The Ti (anode) was grown by magnetron sputtering in an Ar atmosphere of the pressure $5 \times 10^{-3}$ mbar at room temperature. And the Au layer (cathode) was grown through thermal evaporation in the vacuum chamber. Since Ti layer is only 6 nm in thickness, the laser light can pass through it without obvious reflection from the metal layer. The electro-fatigued process of STO crystal was done by sweeping up and down of electric field (0 to 10 kV/cm) with Keithley 2410 source.

**The SHG measurement.** The 45° reflection geometry with fundamental wavelength at 800 nm (150 fs duration at 1 kHz repetition rate) was used here. A half-wave plate was used to rotate the polarization angle of the incident pump pulses and a Glan prism was used to rotated the polarization angle of the output SH pulses. The SH photons, selected by the monochromer, were transformed by a photomultiplier tube and then recorded by a lock-in amplifier (details can be found in Supplementary Figure S1).

**SUPPLEMENTAL INFORMATION**

Supplemental Information can be found online

**ACKNOWLEDGMENTS**

This work is supported by the National Key R&D Program of China (Grant No. 2017YFA0303603, 2016YFA0300701, 2016YFA0401803), the National Natural Science Foundation of China (NSFC; Grant No. 11574316, 11520101002, U1532155), the Key Research Program of Frontier Sciences, CAS (Grant NO. QYZDB-SSW-SLH011), the Innovative Program of Development Foundation of Hefei Center for Physical Science and Technology (Grant No. 2016FXCX002), and the One Thousand Youth Talents Program of China. A portion of this work was performed on the Steady High Magnetic Field Facilities (Ultrafast optical measurement system under superconducting magnet), High Magnetic Field Laboratory, CAS.

**AUTHOR CONTRIBUTIONS**

Z. S., J. S. and Z. Z. conceived the idea. X. X., H. Z. prepared and characterized the samples. X. X. performed SHG measurements. X. X., H. Z., R. Z., L. Y., H. H., L. S., and J. W. performed other characterizations. X. X., Z. S., J. S. and Z. Z. analyzed the data and wrote the manuscript together with discussion with all authors.

**DECLARATION OF INTERESTS**

The authors declare no competing financial interests.

**REFERENCES**

1. Kittel, C. (1976). Introduction to Solid State Physics, Wiley, New York, USA.
2. Sze, S. M., and Ng, K. K. (2007). Physics of Semiconductor Devices 3rd Edition, John Wiley & Sons, New York, USA.
3. Yang, M., Amit, K. C., Garcia-Castro, A. C., Borisov, P., Bousquet, E., Lederman, D., Romero, A. H., and Cen, C. (2017). Room temperature ferroelectricity in fluoroperovskite thin films. Sci Rep. 7, 7182.
4. Misirlioglu, I. B., Okatan, M. B., and Alpay, S. P. (2010). Asymmetric hysteresis loops and

smearing of the dielectric anomaly at the transition temperature due to space charges in ferroelectric thin films. J. Appl. Phys. 108, 034105.

5. Jiang, W., Noman, M., Lu, Y. M., Bain, J. A., Salvador, P. A., and Skowronski, M. (2011). Mobility of oxygen vacancy in $SrTiO_3$ and its implications for oxygen-migration-based resistance switching. J. Appl. Phys. 110, 034509.

6. Cai, H. L., Wu, X. S., and Gao, J. (2009). Effect of oxygen content on structural and transport properties in $SrTiO_{3-x}$ thin films. Chem. Phys. Lett. 467, 313.

7. Trabelsi, H., Bejar, M., Dhahri, E., Sajieddine, M., Khirouni, K., Prezas, R., Melo, B. M. G., Valente, M. A., and Graca, M. P. F. (2017). Effect of oxygen vacancies on $SrTiO_3$ electrical properties. J. Alloy. Compd. 723, 894.

8. Wang, C. C., Lei, C. M., Wang, G. J., Sun, X. H., Li, T., Huang, S. G., Wang, H., and Li, Y. D. (2013). Oxygen-vacancy-related dielectric relaxations in $SrTiO_3$ at high temperatures. J. Appl. Phys. 113, 094103.

9. Brovko, O. O., and Tosatti, E. (2017). Controlling the magnetism of oxygen surface vacancies in $SrTiO_3$ through charging. Phys. Rev. Mater. 1, 044405.

10. Li, Y., Lei, Y., Shen, B. G., and Sun, J. R. (2015). Visible-light-accelerated oxygen vacancy migration in strontium titanate. Sci Rep. 5, 14576.

11. Hanzig, J., Zschornak, M., Hanzig, F., Mehner, E., Stocker, H., Abendroth, B., Roder, C., Talkenberger, A., Schreiber, G., Rafaja, D., Gemming, S., and Meyer, D. C. (2013). Migration-induced field-stabilized polar phase in strontium titanate single crystals at room temperature. Physical Review B 88, 024104.

12. Li, Y. Y., Wang, Q. X., An, M., Li, K., Wehbe, N., Zhang, Q., Dong, S., and Wu, T. (2016). Nanoscale Chemical and Valence Evolution at the Metal/Oxide Interface: A Case Study of $Ti/SrTiO_3$. Adv. Mater. Interfaces 3, 1600201.

13. Fu, Q., and Wagner, T. (2007). Interaction of nanostructured metal overlayers with oxide surfaces. Surf. Sci. Rep. 62, 431.

14. Unoki, H., and Sakudo, T. (1967). Electron spin resonance of $Fe^{3+}$ in $SrTiO_3$ with special reference to 110 degrees K phase transition. J. Phys. Soc. Jpn. 23, 546.

15. Zhong, W., and Vanderbilt, D. (1995). Competing structural instabilities in cubic perovskites. Physical Review Letters, 74, 2587.

16. Bussmannholder, A., Bilz, H., Bauerle, D., and Wagner, D. (1981). A polarizability model for the ferroelectric mode in semiconducting $SrTiO_3$. Z. Phys. B-Condens. Mat. 41, 353.

17. Bloembergen, N., and Pershan, P. S. (1962). Light waves at boundary of nonlinear media. Physical Review 128, 606.

18. Shen, Y. R. (1989). Optical 2nd harmonic-generation at interfaces. Annu. Rev. Phys. Chem. 40, 327.

19. Fiebig, M., Pavlov, V. V., and Pisarev, R. V. (2005). Second-harmonic generation as a tool for studying electronic and magnetic structures of crystals: review. J. Opt. Soc. Am. B-Opt. Phys. 22, 96.

20. Jalan, B., Engel-Herbert, R., Mates, T. E., and Stemmer, S. (2008). Effects of hydrogen anneals on oxygen deficient SrTiO(3-x) single crystals. Applied Physics Letters 93, 052907.

21. Hikita, Y., Kozuka, Y., Susaki, T., Takagi, H., and Hwang, H. Y. (2007). Characterization of the Schottky barrier in $SrRuO_3$/Nb : $SrTiO_3$ junctions. Applied Physics Letters 90, 143507.

22. Robertson, J., and Chen, C. W. (1999). Schottky barrier heights of tantalum oxide, barium strontium titanate, lead titanate, and strontium bismuth tantalate. Applied Physics Letters 74, 1168.

23. Zhong, Z., and Hansmann, P. (2016). Tuning the work function in transition metal oxides and their heterostructures. Physical Review B 93, 235116.

24. Trier F. (2016). Quantum and field effects of oxide heterostructures. doctor thesis, Technical University of Denmark.

25. Jackmana, M. J., Deak, P., Syres, K. L., Adell, J., Thiagarajan, B., Levy, A., Thomas, A. G. (2014). Observation of vacancy-related polaron states at the surface of anatase and rutile TiO2 by high-resolution photoelectron spectroscopy. arXiv 1406. 3385.

26. Tan, H. Q., Zhao, Z., Zhu, W. B., Coker, E. N., Li, B. S., Zheng, M., Yu, W. X., Fan, H. Y., and Sun, Z. C. (2014). Oxygen Vacancy Enhanced Photocatalytic Activity of Pervoskite $SrTiO_3$. ACS Appl. Mater. Interfaces 6, 19184.

27. Petzelt, J., Ostapchuk, T., Gregora, I., Rychetsky, I., Hoffmann-Eifert, S., Pronin, A. V., Yuzyuk, Y., Gorshunov, B. P., Kamba, S., Bovtun, V., Pokorny, J., Savinov, M., Porokhonskyy, V., Rafaja, D., Vanek, P., Almeida, A., Chaves, M. R., Volkov, A. A., Dressel, M., and Waser, R. (2001). Dielectric, infrared, and Raman response of undoped $SrTiO_3$ ceramics: Evidence of polar grain boundaries. Physical Review B, 64, 184111.

28. Jang, H. W., Kumar, A., Denev, S., Biegalski, M. D., Maksymovych, P., Bark, C. W., Nelson, C. T., Folkman, C. M., Baek, S. H., Balke, N., Brooks, C. M., Tenne, D. A., Schlom, D. G., Chen, L. Q., Pan, X. Q., Kalinin, S. V., and Gopalan, V., Eom, C. B. (2010). Ferroelectricity in Strain-Free $SrTiO_3$ Thin Films. Physical Review Letters 104, 197601.

*Supplementary Information*

**Polar Rectification in Electro-Fatigued SrTiO$_3$ Based Junctions**

Xueli Xu, Hui Zhang, Zhicheng Zhong, Ranran zhang, Lihua Yin, Yuping Sun, Haoliang Huang, Yalin Lu, Yi Lu, Chun Zhou, Zongwei Ma, Lei Shen, Junsong Wang, Jiandong Guo, Jirong Sun, Zhigao Sheng

## Table of Contents

## A. The description of Second Harmonic Generation (SHG) measurements.

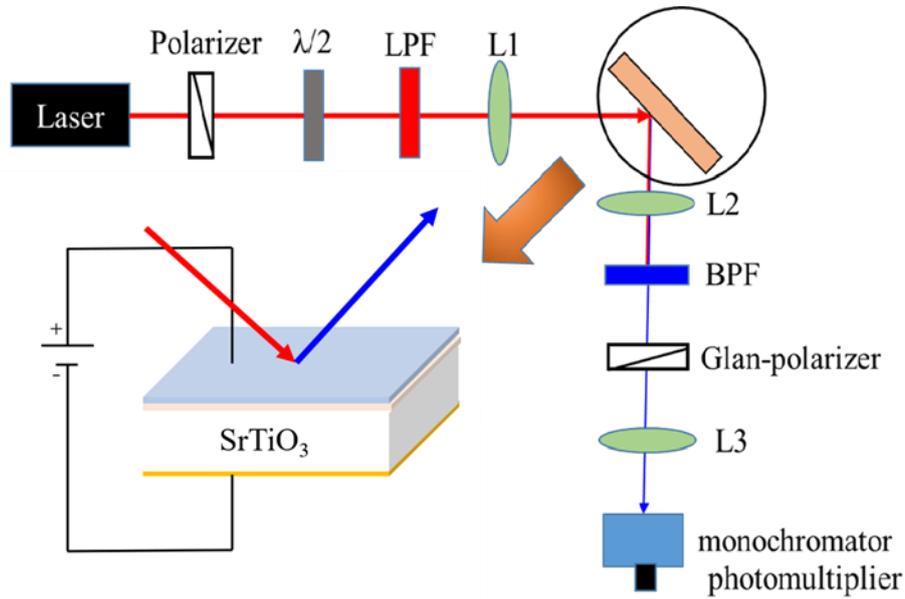

**Figure S1.** Schematic for optical second harmonic generation in the reflection geometry. Focusing lenses are marked by L1, L2, L3. A half-wave plate ($\lambda/2$) is used to rotate the polarization axis of the incident fundamental light. A long-pass filter (LPF) is placed before the sample to filter out residual SHG light from the laser system. A band-pass filter (BPF) is used to block fundamental light from entering the photodetector (PMT). The inset area is the amplification of STO based heterojunction.

As shown in Figure S1, the SHG measurement was carried out in a 45° reflection geometry with fundamental wavelength at 800 nm (150 fs duration at 1 kHz repetition rate). A long-pass filter was used in the incident optical path to ensure that only the fundamental light ($\lambda = 800$ nm) could arrive to the sample, and the short-pass filter in the reflected direction could filter the fundamental light (800 nm) mixed with the SH light (400 nm). A half-wave plate was used to rotate the polarization angle, $\theta$, of the incident pump pulses where $\theta$ is the angle between the polarization axis of the incident light and the *xz*-plane. Glan prism was used to rotated the polarization angle, $\psi$, of the SH pump pulses where $\psi$ is the angle between the polarization axis of the SH light and the *xz*-plane. The SH photons selected by the monochrometer were transformed by photomultiplier tube and then detected by lock-in amplifier.

## B. Band structural analysis of the Ti/SrTiO₃/Au junctions

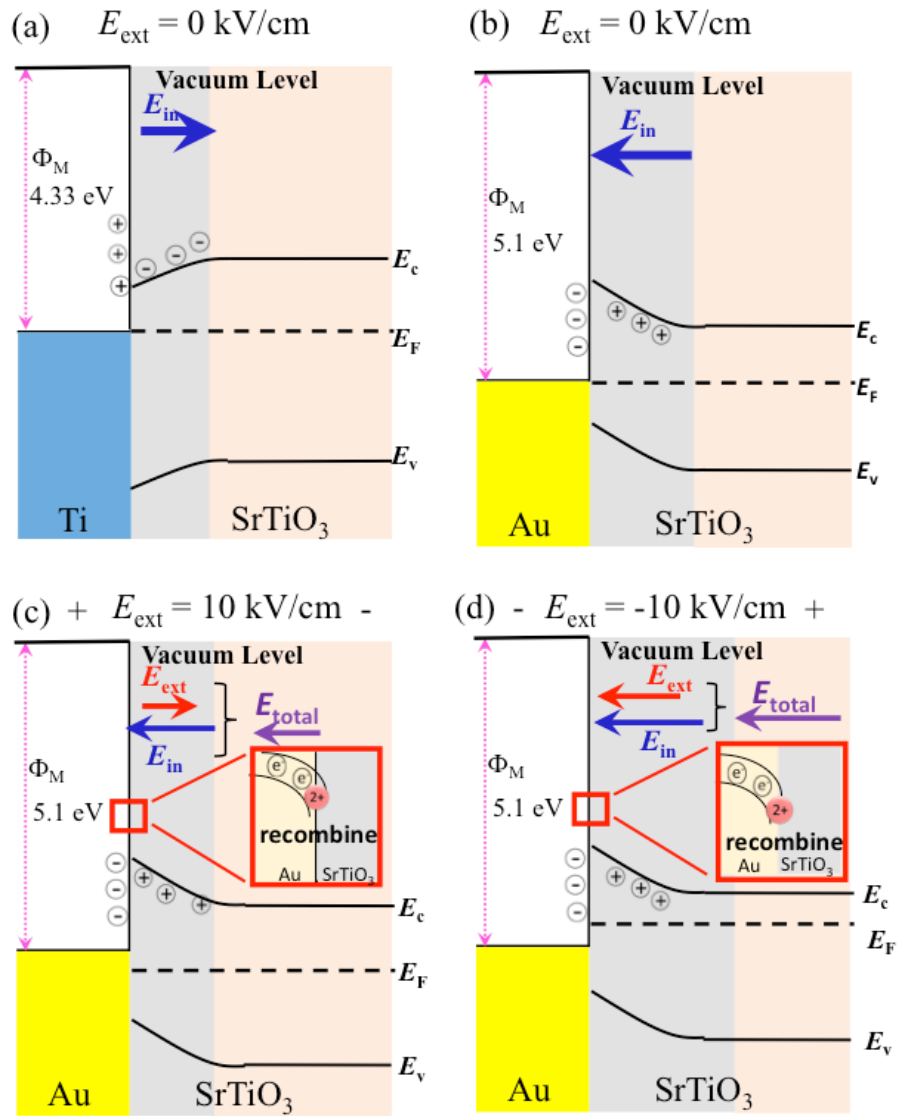

**Figure S2.** The schematic illustration of energy band diagram of (a) Ti/STO Schottky junction at $E_{ext}$ = 0 V/cm and Au /STO Schottky junction for (b) $E_{ext}$ = 0 V/cm, (c) $E_{ext}$ = 10 V/cm and (d) $E_{ext}$ = -10 V/cm.

✧ Simple calculation for the width and the electric field in the depletion layer of Ti/STO junction

In the metal/SrTiO₃ (STO) Schottky junction, when $n_m \gg n_N$ the depletion layer width ($W$) can be written as:

$$W = [\frac{2\varepsilon_s}{q}\left(\frac{n_m+n_N}{n_m n_N}\right)(V_{bi} \pm V)]^{1/2} \approx [\frac{2\varepsilon_s}{q}\left(\frac{1}{n_N}\right)(V_{bi} \pm V)]^{\frac{1}{2}} \quad (S1)$$

where $n_m$ is the carrier concentration of metal and $n_N$ is the oxygen vacancies concentration of STO, $\varepsilon_s$ is the dielectric constant of STO (300 at room temperature), $q$ is the electron charge, $V_{bi}$ is the built in electric voltage of metal/STO Schottky junction, and $V$ is the external electric voltage. For Ti/STO junction, the oxygen vacancies concentration $n_N$ is estimated as $5*10^{17}$/cm$^3$, and the built in voltage $V_{bi}$ = 0.27 V.[1] Accordingly, the $W$ with different applied voltages can be obtained as

1) When $V$ = 0 V, the depletion layer $W_0$ = 133.7 nm

2) When $V$ = 500 V, the depletion layer $W_{500}$ = 5.759 μm

According to Gauss's law, the electric field as a function of position in the depletion layer can be given by:

$$\begin{cases} E_{(x)} = -\frac{qN_d}{\varepsilon_s}(W_0 - W) & 0 < W < W_0 \\ E_{(x)} = 0 & W_0 \leq W \end{cases} \quad (S2)$$

It can be found that the largest value of the electric field ($E_{in}$) is located at the interface and it can be calculated as:

$$E_{x=0} = -\frac{qN_dW_d}{\varepsilon_s} = -\frac{1.6*10^{-19}C \times 5*10^{17}cm^{-3}}{300*8.854\times 10^{-12}CV^{-1}m^{-1}} * 133.7nm = 40.2kV/cm \quad (S3)$$

✦ Simple calculation for the width and the electric field in the depletion layer of Au/STO junction

For the Au/STO junction, the built-in voltage $V_{bi}$ = 0.5 V. Similar to the Ti-STO case, the $W$ with different applied voltages can be obtained as

1) When $V$ = 0 V, the depletion layer $W_0$ = 182.3 nm

2) When $V$ = -500 V ($E$ = 10kV/cm), the depletion layer $W_{-500}$ = 5.759 μm

Also, the electric field in the depletion layer can be calculated based on Eq S2. It is found that the largest value of $E_{in}$ in Au-STO junction is obtained at the interface and it is given by:

$$E_{x=0} = -\frac{qN_dW_d}{\varepsilon_s} = -\frac{1.6*10^{-19}C \times 5*10^{17}cm^{-3}}{300*8.854\times 10^{-12}CV^{-1}m^{-1}} * 182.3nm = 54.9kV/cm \quad (S4)$$

Next, let's discuss why the oxygen vacancies (OVs) and the polarization rectification cannot arise in Au/STO Schottky junction. Figure S2 shows the schematic illustration of energy band diagram for Ti/STO Schottky junction and Au/STO Schottky junction. For Ti/STO Schottky junction, the $E_{in}$ is pointed from Ti to STO as shown in Figure S2a. The OVs will be produced at the interface and be driven into the STO crystal due to the coaction of $E_{in}$ and external electric field ($E_{ext}$). For Au/STO junction, however, the situation is quite different. As shown in Figure S2b, the $E_{in}$ is pointed from STO to Au due to the different band structure. Moreover, it was found that the $E_{in}$ is larger than $E_{ext}$, and hence the movement of extra charges in the junction is dominated by the direction of $E_{in}$. Consequently, even there is OVs can be generated at the Au/STO interface, the OVs will be driven to the Au electrode and then they will be neutralized by the electrons from Au electrode. As a result, no remnant OVs will be produced in the Au/STO junction, let along the polarization. The detail band bending structure of Au/STO junction with different direction of $E_{ext}$ are shown in the Figure S2c and S2d.

Another possible reason for the difficult of the OV producing in Au/STO is the ability of the oxidation. It was found that the Ti is easy to be oxidized than the Au.[2] Hence, it can be expected the Ti is easy to catch oxygen from STO than the Au, i.e., the OVs are easy to be created in Ti/STO junction than in Au/STO with assistance from the coaction of $E_{in}$ and $E_{ext}$.

## C. The XRD results with application of $E_{ext}$ for both Ti/STO and Au/STO junctions

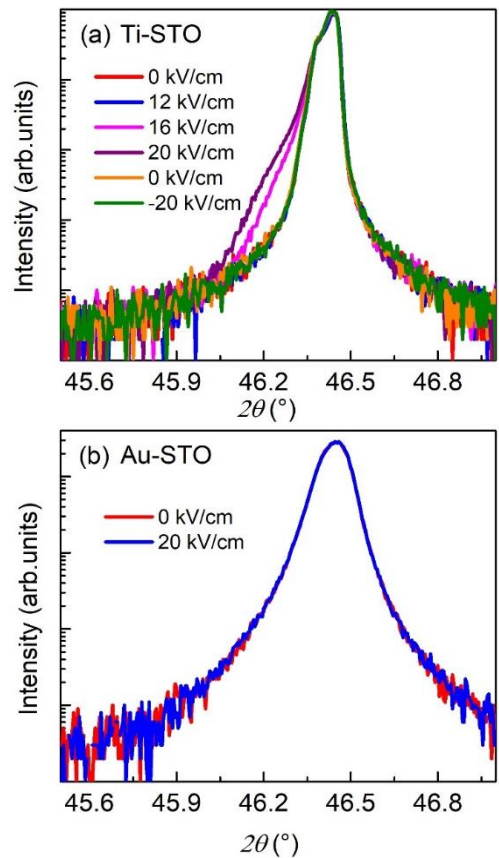

**Figure S3.** The X-ray diffraction (XRD) results of (002) peak for (a) Ti/STO and (b) Au/STO crystal with different $E_{ext}$. The XRD results of (002) peak at different electric field for STO demonstrate that the OVs movement in Ti/STO and Au/STO are different. For Ti/STO, as shown in Figure S3a, the (002) peak is broadened when $E_{ext}$ exceeds 12 kV/cm. However, there is no obvious change when a negative $E_{ext}$ is even up to -20 kV/cm. For the Au/STO junction, it is interesting to find that the crystal lattice remains the same until the $E_{ext}$ reaches 20 kV/cm [the direction of $E_{ext}$ used here is pointed from Au to STO].

## D. Electro-degradation process for the fatigued SrTiO$_3$

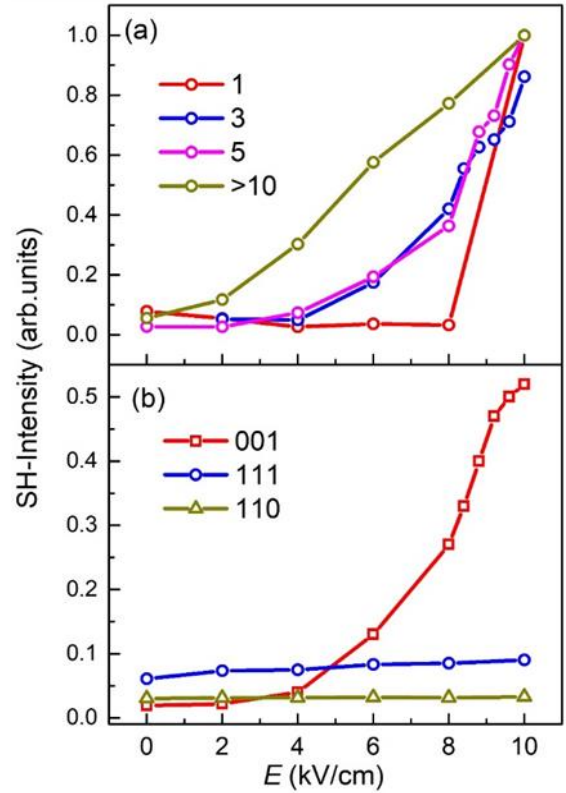

**Figure S4.** (a) The normalized electric field dependence of SH intensity with P$_{in}$-P$_{out}$ geometry measured in Ti/(001)-STO with different times (n = 1, 3, 5 and more than 10 times) of the the electro-fatigued process (sweeping up of $E_{ext}$). (b) The electric field dependence of SH intensity measured for Ti/(001)-, (111)-, (110)-STO junctions. For the electro-fatigued process, the STO single crystals (5*5*0.5 mm$^3$) was covered with a 6 nm-Ti (anode) on the surface and a 100 nm-Au (cathode) on the bottom. We applied the electric voltage up to 500 V (10 kV/cm) at the rate of 2 V/s on STO. And then the 10 kV/cm field was kept on Ti/STO for 30 minutes. For *n* = 1, there is little change of detected SH until $E_{ext}$ exceeds 8 kV/cm, and then there is a sharp increasing of SH polarization when $E_{ext}$ = 10 kV/cm. It implies that the critical electric field ($E_c$) for the creation of OVs is around 8 kV/cm for the Ti/STO interface. As the time of applied electric fields increases, the $E_c$ is decreasing. As shown in the Figure S4a, with sweeping up and down of $E_{ext}$, it become easy to detect the polarization which implies that more and more OVs are created during electro-fatigued process. The fatigued STO

(fSTO) was formed after repeating this process more than ten times. Same electro-fatigued process was done for both Ti/(111)-STO and Ti/(110)-STO junctions. However, as shown in the Figure S4b, the output SH signals for both (111) and (110) cases are close to zero and have no change with increase of $E_{ext}$ up to 10 kV/cm. It implies that the $E_c$ is crystalline direction dependent ant it is much larger for (111) and (110) cases than that for (001) crystal.

E. Calibration of C1s XPS spectra for fatigued and pristine SrTiO$_3$

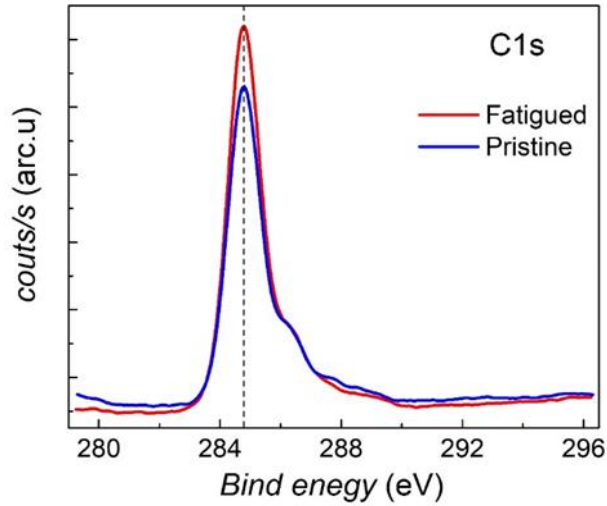

**Figure S5.** The C1s XPS spectra of both pristine STO (pSTO) and fSTO. In maintext, the XPS spectra of both Ti and Sr were measured to identify the existence of OVs. To eliminate the experimental errors of XPS data, the C1s spectra were measured to calibration the Ti and Sr XPS spectral. As shown in the Figure S5, there is almost no difference between fSTO and pSTO, which indicates that the XPS data obtained here are reliable.

## F. Dielectric constant (ε) for both pSTO and fSTO crystals

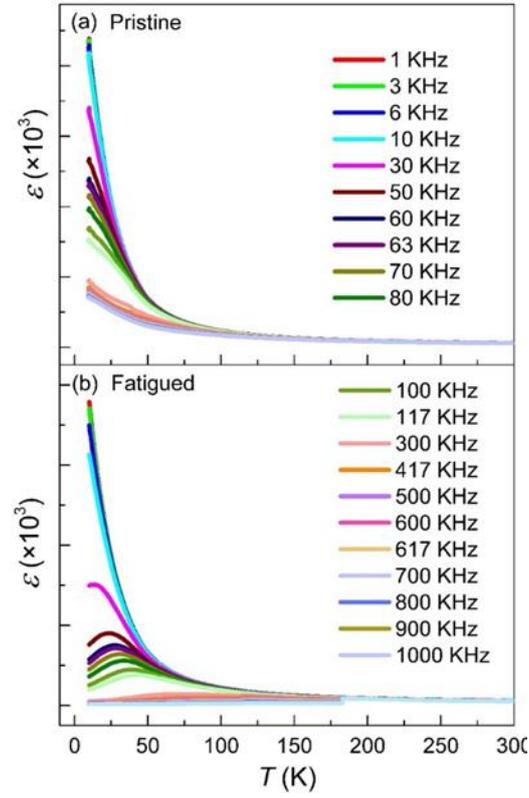

**Figure S6.** The temperature dependence of dielectric constant ($\varepsilon$) with different frequency for both (a) pSTO and (b) fSTO crystals respectively. For pSTO crystal as shown in Figure. S6a, the $\varepsilon$ increases with the temperatures decrease from 300 K down to 10 K. Moreover, the $\varepsilon$ decreases with increase of the frequencies of the ac electric field ($f$). These features are consistent with the previous reports.[3] For fSTO crystal as shown in Figure S6b, the $\varepsilon$ also increases with the temperature decrease from 300 K down to 10 K. However, with increasing of $f$ up to 30 kHz, there is a broad dielectric peak emerges at low temperatures. Furthermore, the peak position shifts to high temperatures when the $f$ increases. One point should be noted that such dielectric peak observed here is not as sharp as that in the STO with ferroelectric state,[4] and its temperature and frequency dependent behaviors are similar to those of relaxed ferroelectricity.[5-6] Such relaxed behavior might be resulting from the contribution of separated OVs-electron pairs in the depletion layer of Ti/STO junction. The polarized top layer (~ 5 um) mixed with un-polarized bottom layer (~ 495 um) could produces a relaxed ferroelectric behavior as shown in the Figure S6b.

### G. The estimation of OV concentration at the surface of fatigued STO

The existence of OVs plays an important role in the realization of both pre-ferroelectric state and polar rectification property in Ti/fSTO heterojunctions. It was reported previously, the STO with different OV concentration could hold phase transition from insulator to metal.[7] Accordingly, it is important to estimate the oxygen vacancy concentration in the fSTO crystal used here.

Firstly, the color of the fSTO and pSTO crystal were checked. It was found that there is no difference between two kinds of crystal. Both of them show colorless, even for the surface of fSTO sample. It implies that the OV concentration is low and the fSTO is still an insulator because the surficial color would be blackened if STO become conductive.[8,9]

Secondly, the electronic conductance of fSTO crystal were measured roughly. With an ohmmeter, the resistance of the fSTO crystal surface was measured. It was found that the resistance (R) of fSTO surface is over rang of the ohmmeter, *i.e.*, R > 200 MΩ, indicating a conductance $\sigma_{dc} < 5\times10^{-8}(\Omega cm)^{-1}$. In 1993, P. Calvani et al. have obtained that there is a corresponded relationship between conductance and the carrier concentration (N) as shown in the following Table 1.[10] Accordingly, it can be found that the carrier concentration of fSTO should be lower than $5\times10^{10}$ cm$^{-3}$.

**TABLE 1**. The values of room-temperature dc conductivity, which are determined by four wire measurements, the carrier concentration, and the $\sigma_{dc}$ of different samples.[10]

| Sample | $\sigma_{dc}$ (Ωcm)$^{-1}$ | N (cm$^{-3}$) |
|---|---|---|
| N | $1.1\times10^{-8}$ | $10^1$ |
| M | $\times10^{-8}$ | $5\times10^{10}$ |
| B | 0.25 | $2.5 \times 10^{17}$ |
| P | 2.2 | $2 \times 10^{18}$ |

At present stage, it is difficult for us to obtain the exact OV concentration of fSTO crystal and it will be done in our next study in near future.

**Reference**

1. Uedono, A., Shimayama, K., Kiyohara, M., Chen, Z. Q., and Yamabe, K. (2002). Study of oxygen vacancies in $SrTiO_3$ by positron annihilation. J. Appl. Phys. 92, 2697.
2. Fu, Q., and Wagner, T. (2007). Interaction of nanostructured metal overlayers with oxide surfaces. Surf. Sci. Rep. 62, 431.
3. Weaver, H. E. (1959). Dielectric properties of single crystals of $SrTiO_3$ at low temperatures. J. Phys. Chem. Solids 11, 274.
4. Haeni, J. H., Irvin, P., Chang, W., Uecker, R., Reiche, P., Li, Y. L., Choudhury, S., Tian, W., Hawley, M. E., Craigo, B., Tagantsev, A. K., Pan, X. Q., Streiffer, S. K., Chen, L. Q., Kirchoefer, S. W., Levy, J., and Schlom, D. G. (2004). Room-temperature ferroelectricity in strained $SrTiO_3$. Nature 430, 758.
5. Jang, H. W., Kumar, A., Denev, S., Biegalski, M. D., Maksymovych, P., Bark, C. W., Nelson, C. T., Folkman, C. M., Baek, S. H., Balke, N., Brooks, C. M., Tenne, D. A., Schlom, D. G., Chen, L. Q., Pan, X. Q., Kalinin, S. V., Gopalan, V., and Eom, C. B. (2010). Ferroelectricity in Strain-Free $SrTiO_3$ Thin Films. Phys. Rev. Lett. 104, 4, 197601.
6. Biegalski, M. D., Jia, Y., Schlom, D. G., Trolier-McKinstry S., Streiffer S. K., Sherman V., Uecker R., Reiche P. (2006). Relaxor ferroelectricity in strained epitaxial SrTiO3 thin films on DyScO3 substrates. Appl. Phys. Lett. 88, 192907.
7. Schooley, J. F., Hosler, W. R., and Cohen M. L. (1964). Superconductivity in Semiconducting $SrTiO_3$. Phys. Rev. Lett. 12, 474.
8. Noll F., Munch W., Denk I., and Maier J. (1996). $SrTiO_3$ as a prototype of a mixed conductor Conductivities, oxygen diffusion and boundary effects. Solid State Ion. 86-8, 711.
9. Potzger, K., Osten, J., Levin, A. A., Shalimov, A., Talut, G., Reuther, H., Arpaco, S., Burger, D., Schmidt, H., Nestler, T., and Meyer, D. C. (2011). Defect-induced ferromagnetism in crystalline $SrTiO_3$. J. Magn. Magn. Mater. 323, 1551.
10. Calvani, P., Capizzi, M., Donato, F., Lupi, S., Maselli, P., and Peschiaroli, D. (1993). Observation of a midinfrared band in $SrTiO_{3-y}$. Phys. Rev. B 47, 8917.